# Instrument-Level Validation of Upper Sideband Syntonization Concept

Volodymyr Kudriashov[1,†], Manuel Martin-Neira[2]

[1]*Serco Nederland B.V., Kapteynstraat 1, 2201 BB Noordwijk, Volodymyr.V.Kudriashov@GMail.com*
[2]*European Space Research and Technology Centre (ESTEC), Keplerlaan 1, 2201 AZ Noordwijk, Manuel.Martin-Neira@ESA.int*

Upper Sideband Syntonization (USBS) is a two-way frequency-transfer concept proposed by the European Space Agency for distributed space radio interferometric systems. Previous investigations validated the concept primarily through frequency-stability and phase-transfer measurements.

This work demonstrates the first end-to-end instrument-level validation of the USBS concept through coherent W-band cross-correlation measurements and image reconstruction.

A P=2 USBS breadboard was integrated with a W-band radio interferometer developed for this test. The synchronized local oscillators (LOs) were evaluated through long-term cross-correlation measurements and coherent imaging of a point-like noise source. The results were compared with those obtained using a conventional common LO as well as two independent free-running LOs.

The USBS-derived LOs maintained coherence over 500 s of cumulative effective integration time acquired over approximately five days of elapsed time and preserved inter-session phase stability over measurements spanning 40 days. The reconstructed image is nearly indistinguishable from that obtained with a common LO, whereas independent free-running LOs lead to severe coherence degradation, image distortion, and approximately 38 dB loss of peak response.

The results demonstrate that USBS preserves the interferometric coherence required for W-band imaging and present the first end-to-end instrument-level validation of the USBS syntonization concept for future distributed space radio interferometric missions.

## 1. Introduction

Phase-coherent operation of spatially separated radio interferometric receivers requires either sufficiently stable independent frequency references or the transfer of a common phase/frequency reference between the receiving elements. This work experimentally evaluates the latter approach.

The Upper Sideband Syntonization (USBS) concept is a two-way frequency-transfer technique proposed in [Martin Neira 2019, 2022, 2023] to establish coherence between spatially separated sensors. It targets the relative frequency and phase stabilities rather than absolute ones.

Previous USBS P=1 demonstration experiments at frequencies up to 100 GHz [Kudriashov, 2021a] demonstrated an Allan deviation of $1\text{x}10^{-14}/\tau$ over 10 ms < $\tau$ < 1,000 s between the two derived LOs (using photodetectors and commercially affordable Mach-Zehnder modulators). The experiments further demonstrated an approximately white-phase-noise behavior of the differential LO phase, with a standard deviation of approximately 0.5° at 100 GHz over more than 3 h [Kudriashov, 2021a]. The experiments also demonstrated both a robustness against ±3°C differential temperature variation between the two satellite-representative breadboards over a 4.5 h orbital period and operation with inter-satellite link asymmetry up to 20 s [Kudriashov, 2021a]. The demonstrated performance also permits sufficient coherence at substantially higher observing frequencies [Rogers, 1981], potentially relaxing the 20% instrument sensitivity-loss budget.

The performance of the basic USBS P=1 architecture depends on the second-harmonic rejection of the employed mixer, imposing a constraint when the two spacecraft operate at identical LO frequencies [Kudriashov, 2021a, 2021b]. To overcome this limitation, the advanced P=2 architecture [Martin Neira, 2023] was developed and experimentally validated in [Kudriashov, 2024, 2025], demonstrating coherent operation up to 15 min at frequency offsets approaching zero. Extensions to synchronization of distributed systems comprising N satellites are discussed in [Martin-Neira, 2026].
These previous experiments characterized USBS primarily in terms of frequency stability and phase-transfer performance. For radio interferometry, however, the relevant instrument-level requirement is preservation of the complex coherence required to recover sky signals and reconstruct an image.
Here, USBS P=2 architecture is evaluated against a common LO and two independent free-running LOs, first through correlation measurements and subsequently through coherent image reconstruction. Power loss, coherence and phase consistency are used to assess the instrument performance. Such distributed phase-coherent architectures are of potential interest for space-based radio astronomy interferometers [Kudriashov, 2019, 2021c].

†Corresponding author.

## 2. Testbed

The testbed consists of the LO breadboard, W-band radio interferometer and laboratory hardware units. The experimental evaluation comprises cross-correlation measurements followed by coherent image formation using the same interferometer.

### 2.1. *Local Oscillators*

The USBS P=2 demonstrator (Fig. 1) detailed in [Kudriashov, 2024] consists of two free-running core oscillators A and B (A`, B` denote their Doppler shifted time delayed copies out of the communications Inter-Satellite Link, ISL), two frequency doublers, four microwave mixers and optoelectronics simulating the ISL. The setup outputs the two synchronized LO signals to the W-band interferometer. Allan deviation between the two 46.3 GHz LOs derived in [Kudriashov, 2024] is approximately $1\times10-14/\tau$.
Since output power levels are insufficient to directly drive the interferometer receivers at LO/2 frequency (46.3 GHz), both outputs are amplified to the required +1 to +2.5 dBm level for this test.

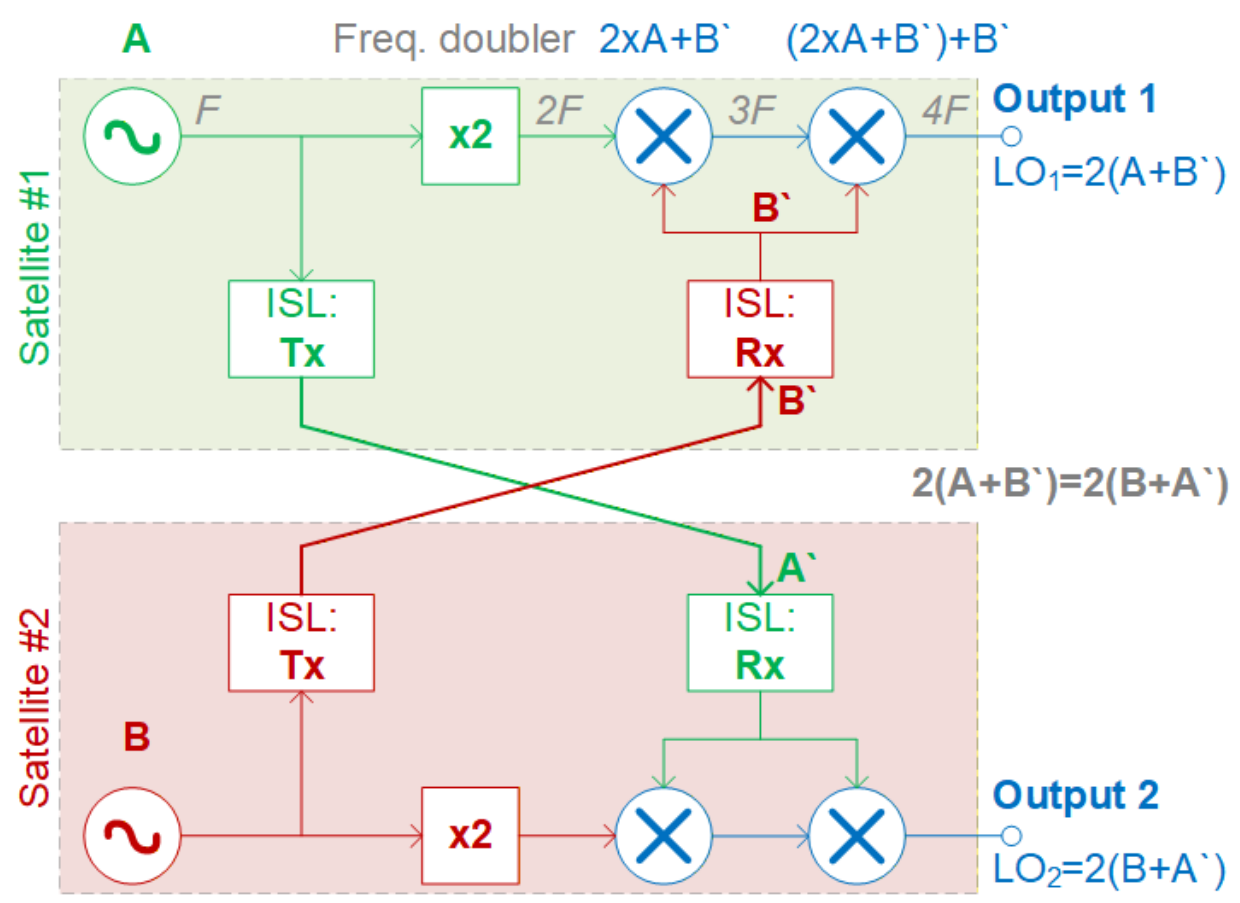

Fig. 1. A simplified block diagram of the USBS P=2 setup [Kudriashov, 2024].

### 2.2. *W-band Interferometer*

The radio interferometer comprises two superheterodyne receivers operating in single sideband mode, an oscilloscope sampling the receiver intermediate-frequency (IF) outputs, and a computer performing signal processing in MATLAB. The test target consists of a noise source and microwave amplifiers.
The receivers originate from hardware delivered by Thomson-CSF (now Thales) to the European Space Agency under the 1996 RADET project. These total-power receivers, originally designed for an interferometric architecture in which phase was not measured, were modified as follows. The input "sky" bandwidth was constrained to enable upper sideband operation. The built-in LOs were disconnected, and external LO-drive ports were added by modifying the microstrip lines and bond wires, IF amplifiers have been added. The required external LO-drive level and frequency were characterized in the ESTEC Microwave Laboratory. The LO-circuitry has been upgraded to permit operation with an external LO-drive at V-band.

The input bandpass filters have a 3 dB passband of 92.65–93.55 GHz. An effective bandwidth of 1 GHz is used in this work, constrained by the performance of the filters, oscilloscopes, and noise generators. The external LO-drive frequency is $F_{LO}/2$=46.3 GHz. Noise figure is 10 dB.

### 2.3. *Imaging Radio Interferometer*

For near-field indoor imaging, antennas with a half-power beamwidth of 32° were used. The experiment was performed indoors using a single artificial noise source.
For aperture synthesis, antennas were moved using Thorlabs Compact Motorized Translation Stages MTS50/M-Z8 and Kinesis software. The stages were homed before each sequence of 22 antenna positions. The antennas were translated stepwise by 0.5 centre wavelength (1.62 mm) between acquisitions. Translation speed and acceleration were minimized, and a settling interval was introduced after each movement before data acquisition. The baseline between the two receiving antennas remained constant while the antenna pair was translated relative to the emitter. The MTS50/M-Z8 positioning error (step error <3 µm; total error <64 µm), projected onto the target direction by sin(8.6°)≈0.15, corresponds to approximately 1° of carrier phase and is therefore negligible.
In the common LO configuration, the LO subsystem (including the amplifier, divider, and semi-rigid receiver-feed cables) was mounted on the same translation stage as the receivers, thereby avoiding phase changes caused by cable flexure. Therefore, the observed differential phase variations are expected to be dominated by the LOs.
For the imaging measurements, a Keysight UXR0402A Infiniium oscilloscope (two channels, 40 GHz, 256 GSa/s, 10 bit) was used. The oscilloscope samples a 1.8 GHz IF bandwidth at 4 GSa/s. Acquisition is controlled by MATLAB and triggered at 4 Hz using the trigger generator listed in Table 1. The digitized signals are processed in MATLAB.

## 3. Correlator level test

Whereas previous studies validated USBS at the LO level, the present experiment evaluates its performance at the instrument-level. The objective is to determine whether the synchronization performance is preserved through the complete W-band receiver and correlator chain and ultimately enables coherent image formation.

### 3.1. *Measurement Principle*

We are using a test target emitter in front of receivers. The cross-correlation of the receiver output signals is

$$R(\tau,T)=\frac{1}{T}\int_0^T s_1(t)s_2^*(t-\tau)dt. \tag{1}$$

where $s_1(t)$ and $s_2(t)$ are digitized receiver-output signals which therefore include receiver thermal noise, phase noise, and ADC effects, $\tau$ denotes a relative time delay between the signals, $T$ is the integration (averaging) time, and * denotes complex conjugation. The real-valued signals out of the oscilloscope channels are converted to analytic signals using the Hilbert transform and subsequently processed by a frequency-domain FX correlator in MATLAB. The correlator level test is dedicated to measuring both absolute value and phase of the cross-correlation as functions of integration time.

### 3.2. *Test set-up*

A test target signal is fed into receivers using a directional coupler with attenuators (Fig. 2, Table 1) to equalize the test signal levels. The SNR is approximately two times (6 dB) at IF outputs of both receivers.

Table 1 – Main components of the correlator level test setup

| **Test target** | Microwave Amplifier | *Quinstar* | *QPN-94023025* |
|---|---|---|---|
| | Calibration Noise Source | *NOISE COM* | *NC5110* |
| **Receivers** | Superheterodyne | Thomson-CSF | *RADET W-band receiver* |
| **LO source** | $LO_1$ = VNA (10 MHz master) | *Keysight* | *PNA N5227B* |
| | $LO_2$ = RF Generator (10 MHz slave) | *Agilent* | *83650B* |
| **Data acquisition** | Oscilloscope | *Keysight* | *DSOS404A (later MSOS804A)* |
| | Oscilloscope trigger generator | *Agilent* | *33120A* |
| | Software | *MATLAB* | *instrument control toolbox* |

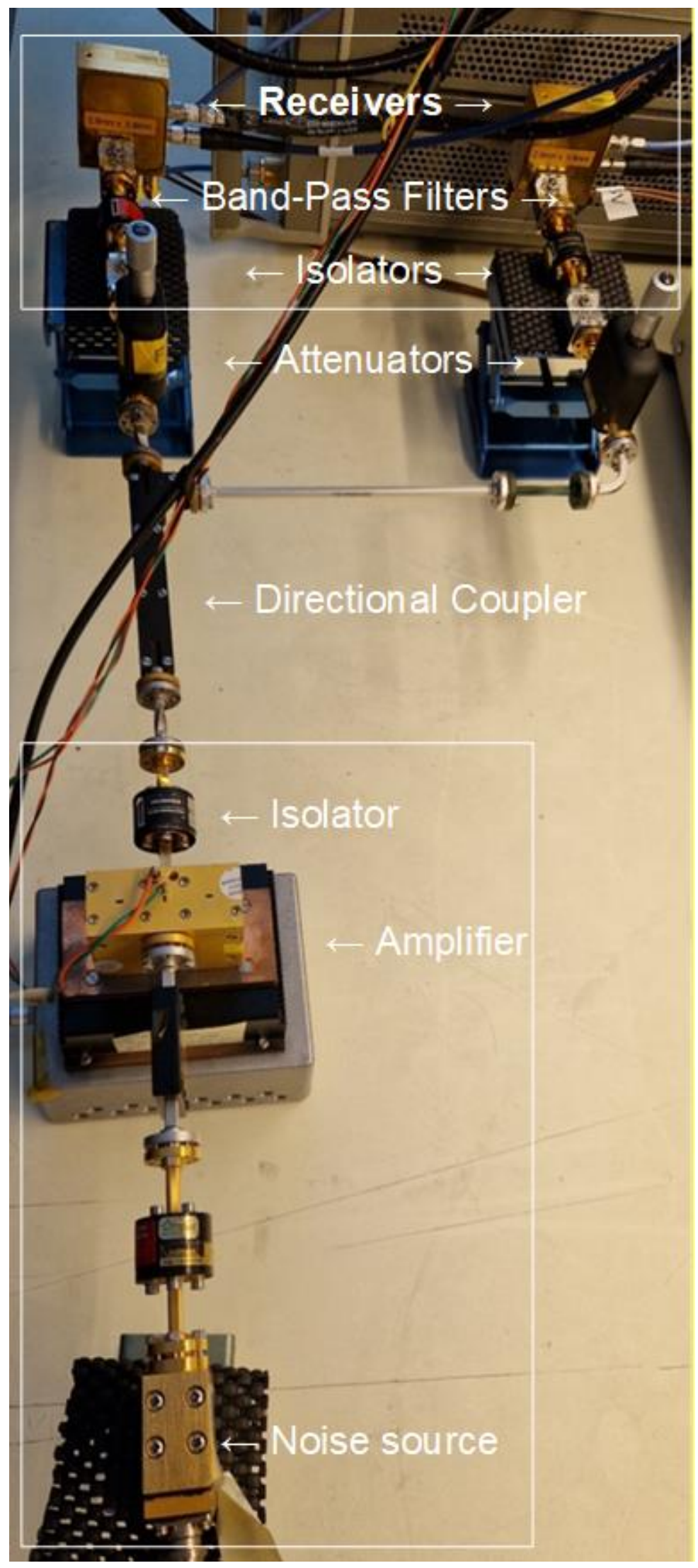

Fig. 2. A test target connected to the two receivers.

Coherent integration was evaluated for three LO configurations: (i) a common LO driving both receivers, (ii) two LOs derived from the P=2 USBS breadboard, and (iii) two independent free-running LOs (Fig. 3).

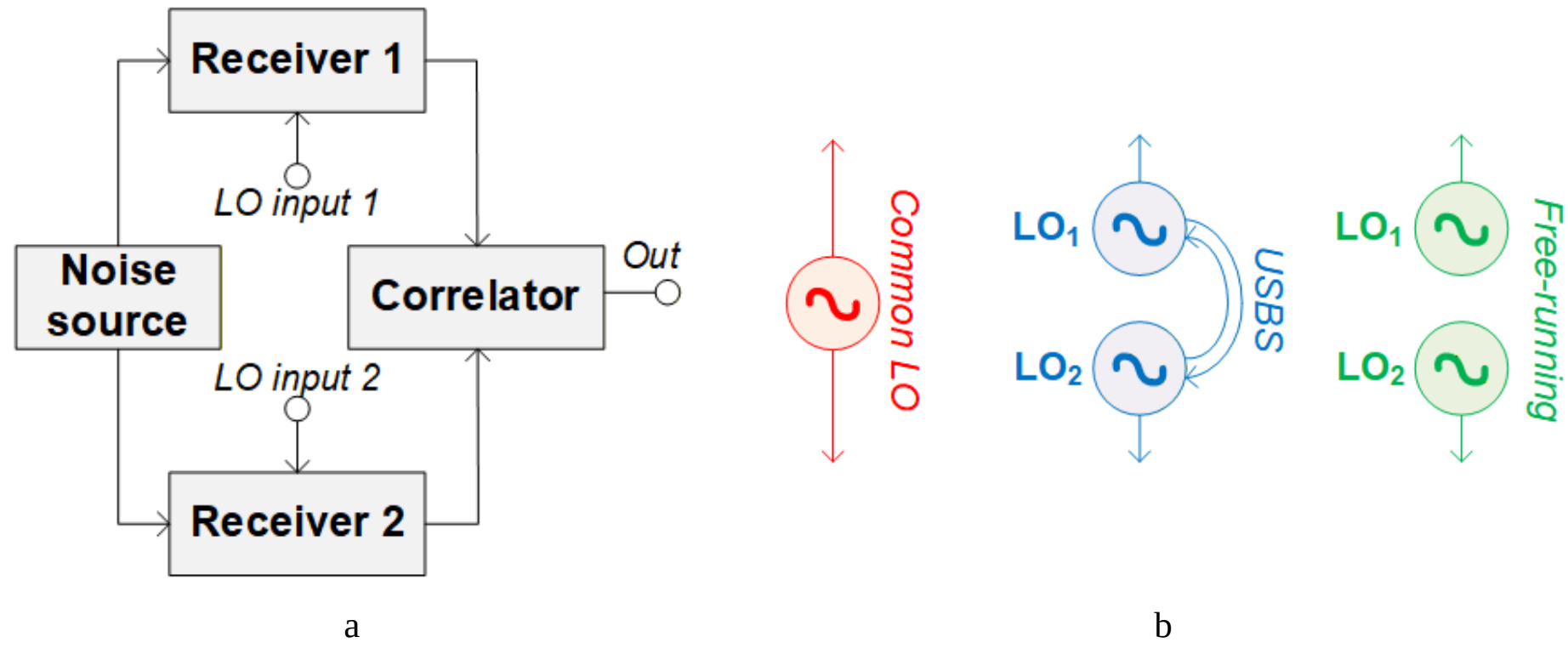

Fig. 3. (a) Block diagram of the cross-correlation measurement. (b) Three LO configurations: common LO (red), two LOs derived from the P=2 USBS setup (blue; main test case), and two independent free-running LOs (green).

### 3.3. *Measurement Results: Correlator Level*

Measurement results are shown in Fig. 4. Red, blue and green curves correspond to block diagrams in Fig. 3b. USBS enables coherent W-band operation up to 500 s of cumulative effective integration acquired over approximately five days of elapsed time. Results are normalized on own maximum values so to avoid a bias by a change in LO drive between three LO configurations under the test.

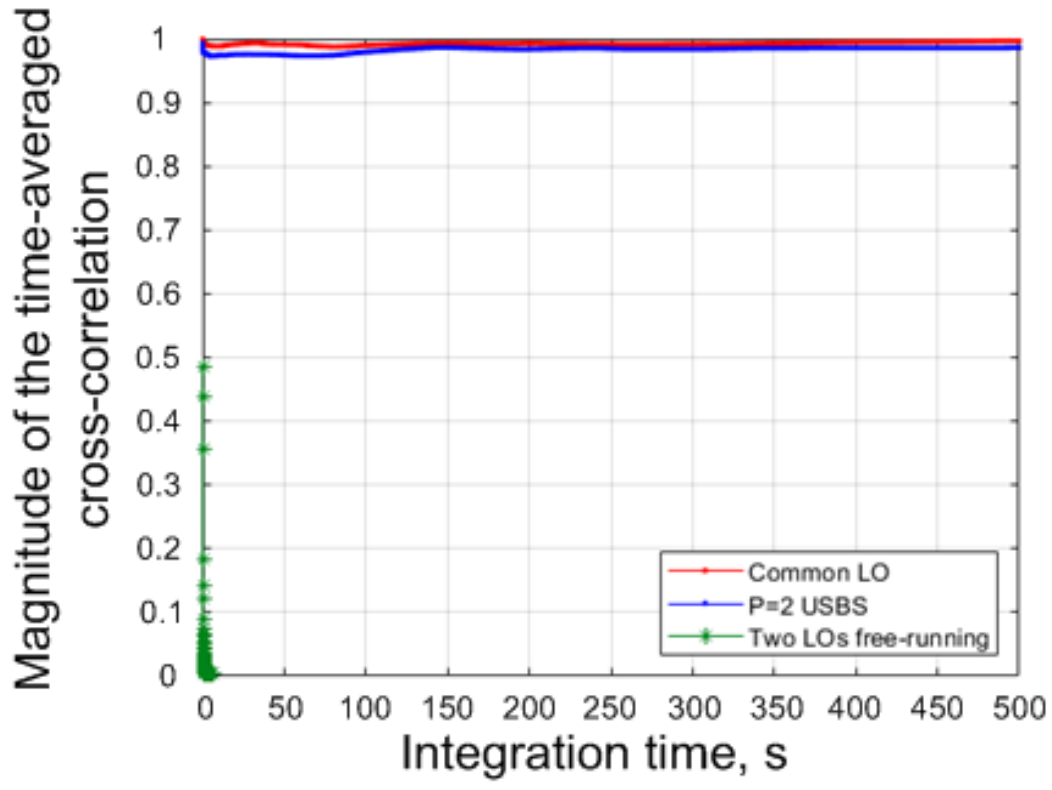

Fig. 4. Absolute value of the cross-correlation measured up to 500 s of integration time, as acquired over a longer elapsed time. Red: common LO; blue: P=2 USBS; green with markers: two independent free-running LOs.

Three timescales are distinguished throughout this work. Each oscilloscope acquisition contains approximately $T_{Acq.}$≈10 ms of continuously acquired data. The cumulative effective integration time $T_{Eff}=\Sigma_k(T_{Acq.,k})=500\ s$ is achieved by processing multiple oscilloscope acquisitions. Because each acquisition is followed by approximately 10 s of data transfer time to the MATLAB computer (Fig. 5a), an elapsed time it takes to accumulate each $T_{Eff}$ is approximately five days.

Although data acquisition was intermittent, the RF and LO hardware remained continuously operating during each multi-day measurement cycle $T_{Elaps}$. Hence, we conclude the setup is capable of long-term coherent operation, Fig. 5b. For measurements unaffected by known setup anomalies, the cross-correlation magnitude remained above 0.94 (solid curves). Measurements affected by amplifier failure, temperature excursions, or additional vibration are shown as dashed curves indicating such would not be present in a nominal operation scenario.

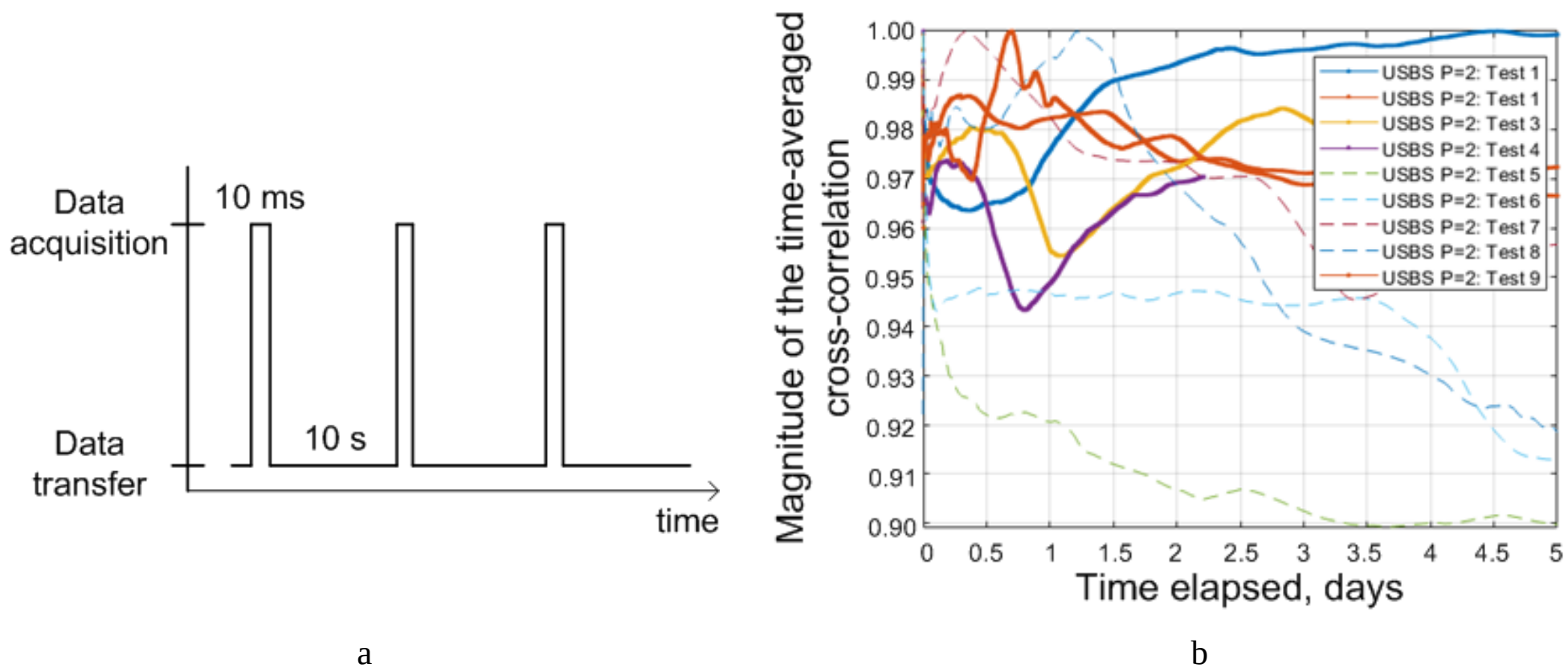

Fig. 5. Oscilloscope data duty cycle (a) and the absolute value of the cross-correlation with respect to the elapsed time

Despite failures in an amplifier (measurement five) and a cooling fan (measurements six, seven), the test setup preserved inter-session phase stability over measurements spanning 40 days, Fig. 6. No phase offset was estimated or removed between the individual measurement sessions. The spike at the beginning of the measurement seven is a measurement artefact caused by a setup fault (a cooling fan failure introduced mechanical vibration and a temperature change in one LO branch).

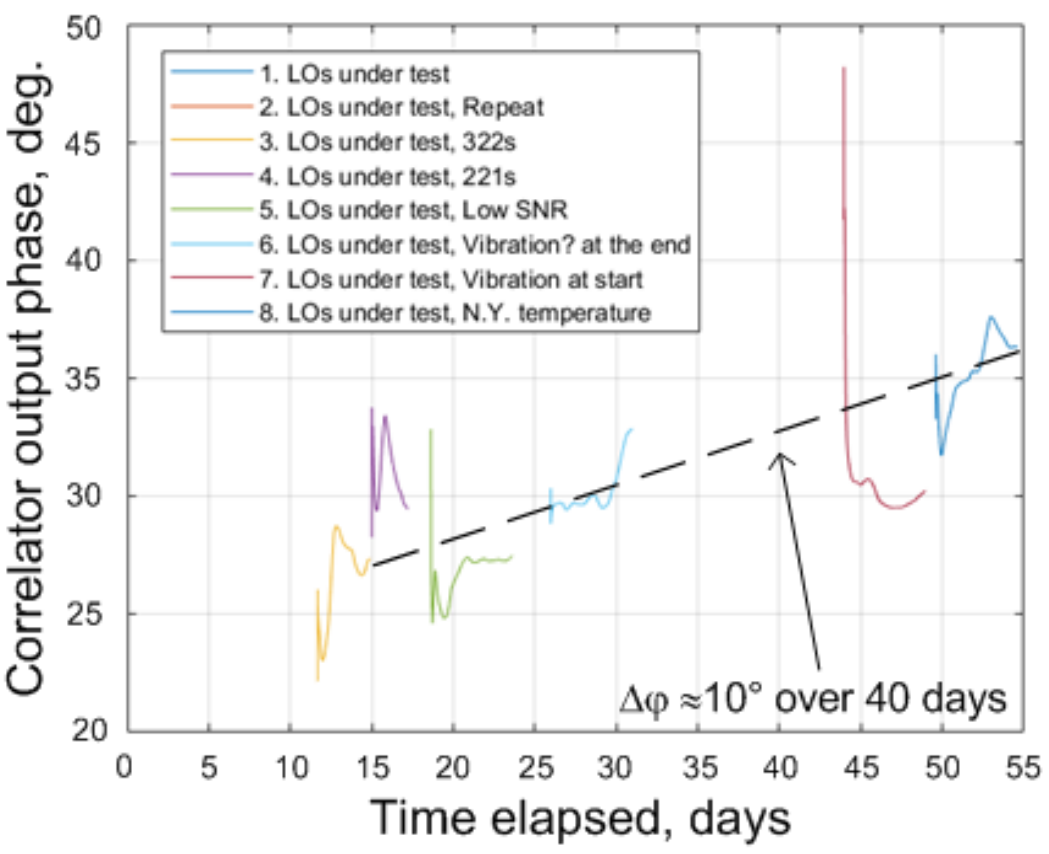

Fig. 6. Phase of cross-correlation in tests 3-8 with respect to the elapsed time. A change of the inter-session phase over a 40-day interval (between measurements 3 and 8) is approximately 10°.

The instrument-level test confirms that P=2 USBS preserves W-band coherence. In addition, the measurements demonstrate inter-session repeatability of the total RF phase over a month-scale interval.

### 3.4. *Mission Implications*

The observed inter-session phase stability suggests that the transferred reference may not dominate the calibration cadence in a space-VLBI instrument. The achievable cadence will also depend on receiver stability, baseline knowledge, thermal variation, and the selected calibration architecture.

## 4. Image Validation Test

The correlator level measurements verify that the synchronization subsystem preserves the complex cross-correlation over long cumulative effective integration times. The next step is therefore to evaluate whether this preservation translates into successful coherent image reconstruction.

### 4.1. *Imaging Approach*

The image is reconstructed by coherently backprojecting the measured complex cross-correlation values over the image grid following [Kudryashov, 2013, 2012]. For a trial pixel located at $\boldsymbol{r}$, the image is

$$I(\boldsymbol{r}) = \sum_{a=1}^{A} C_a(\tau_r)\, exp(-j\, 2\pi f_C\, \tau_r)\,, \tag{2}$$

where $C_a$ is the measured complex cross-correlation for the a-th antenna position and $C_a(\tau_r)$ is its delay bin corresponding to the pixel location $\boldsymbol{r}$; $f_C$ is the centre "sky" frequency; $\tau_r = (R_{1,a} - R_{2,a})/c$ is propagation time delay from location $\boldsymbol{r}$, $R_{1,a}$ and $R_{2,a}$ are the Euclidean distances from trial location $\boldsymbol{r}$ to receivers one and two, respectively, at aperture position $a$, and $c$ is the speed of light. The exponential term compensates the geometric differential delay for each interferometer position such that contributions originating from the true source location are added coherently. The same reconstruction operator and all processing parameters were applied unchanged to the three LO configurations; only the LO architecture was changed.

### 4.2. *Measurement Results: Image Validation Test*

Image-validation test setup is shown in Fig. 7. The emitter is located at a range of 1 m and a cross-range of -15 cm. Due to failure of RF generator, the common-LO is derived from PNA (Table 1), system amplifier (Keysight 83050A, 2-50 GHz) and power divider (Anritsu V240C, DC-65 GHz).

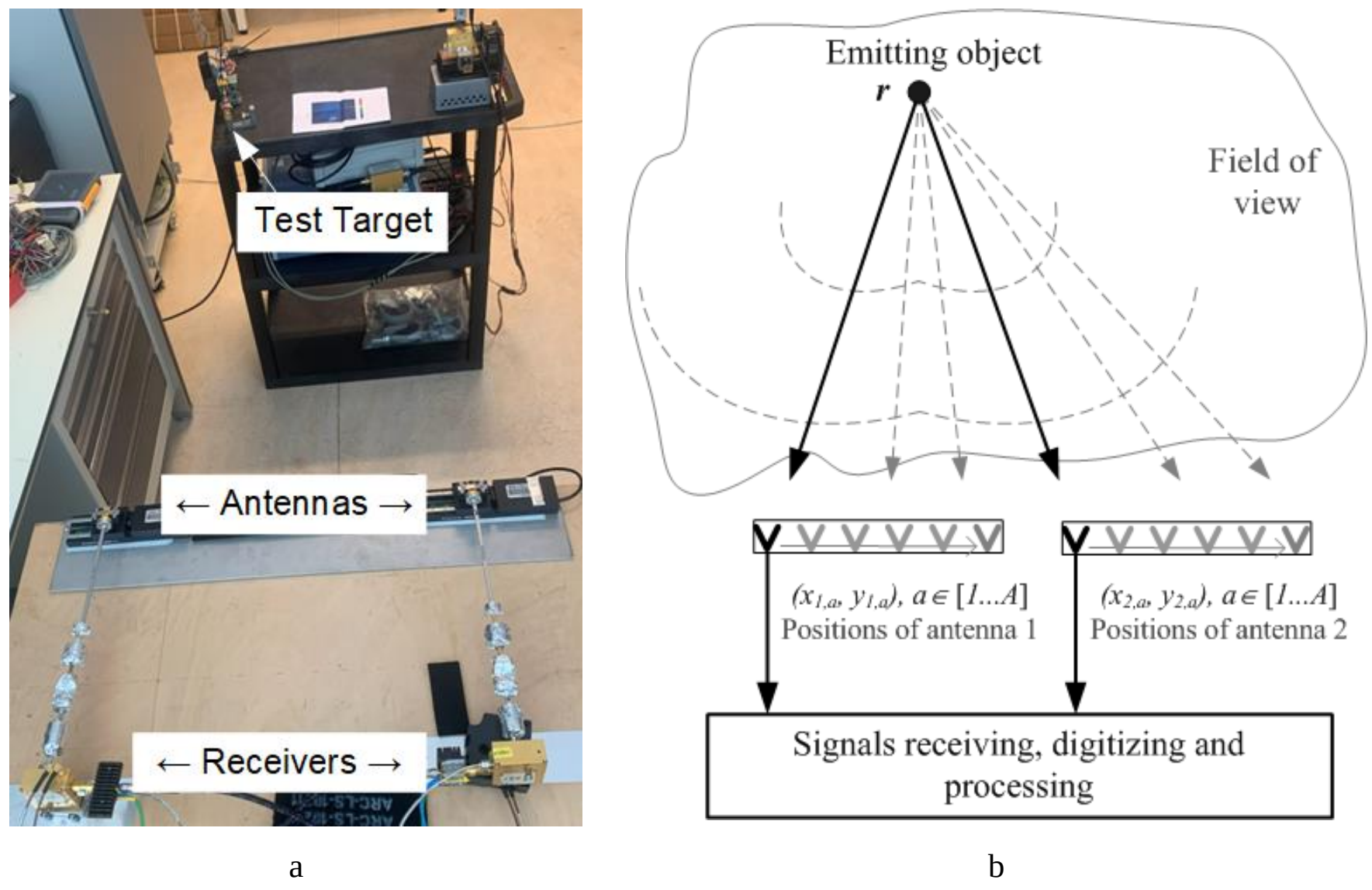

Fig. 7. Image validation test setup (a) and image-reconstruction plane (b) adapted from [14].

Images reconstructed for the three LO configurations are shown in Fig. 8. The emitter images reconstructed with the radio interferometer driven by either the common LO or the two LOs derived from the P=2 USBS demonstrator are very similar, whereas two free-running LOs substantially degrade the reconstructed image. This demonstrates successful image reconstruction with USBS-derived LOs.

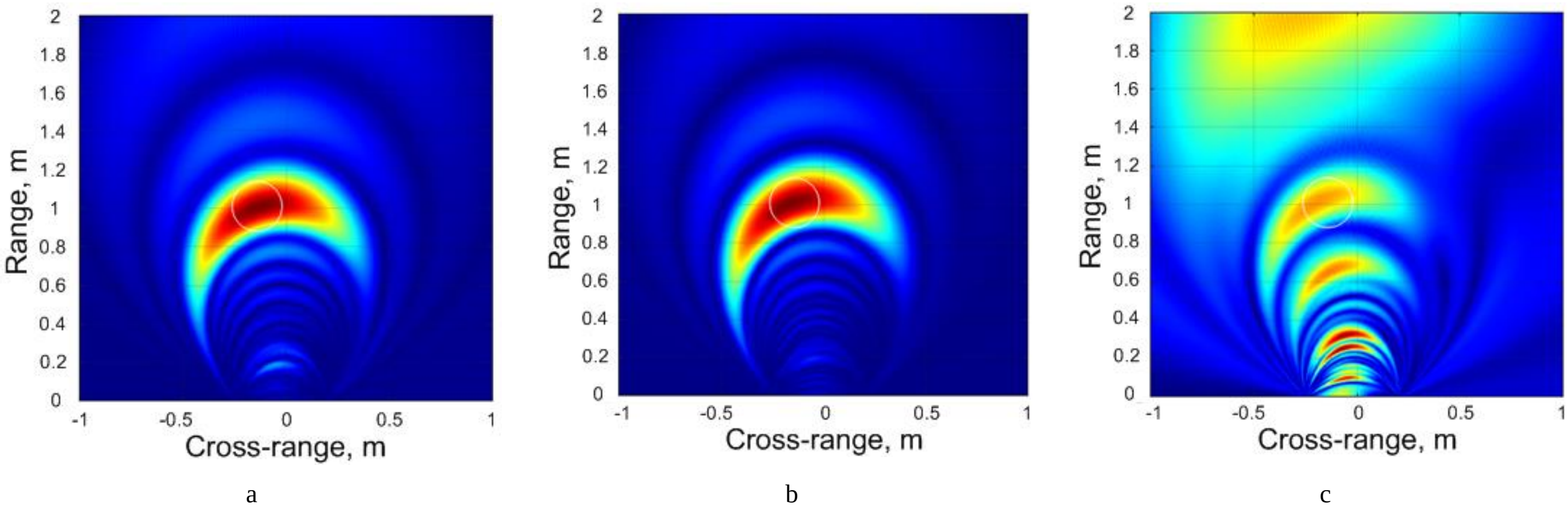

Fig. 8. Images reconstructed using (a) Common LO, (b) USBS-derived LOs and (c) two independent free-running LOs. The true emitter position (range 1 m, cross-range -0.15 m) is indicated by the centre of the white circle. To facilitate comparison of image morphology, each image is normalized to its own maximum.

Independent free-running LOs produce random inter-channel phase variations (an example realization is shown in Fig. 9c) that prevent coherent accumulation across antenna positions. Consequently, the emitter peak is significantly reduced, displaced, and accompanied by elevated sidelobes in Fig. 8c. Because the free-running-LO image is severely corrupted, quantitative comparison is restricted to the magnitude at the known emitter location.

The magnitude losses reported below and in Table 2 are calculated from the not normalized reconstructed images; normalization in Fig. 8 is applied only for visualization of image morphology. The image-magnitude degradation is defined as

$$L = 20\, log_{10} \frac{|I_{CommonLO}(\boldsymbol{r_0})|}{|I_{Test}(\boldsymbol{r_0})|}, \quad (3)$$

where $\boldsymbol{r_0}$ is the known emitter location in the image plane, $I_{CommonLO}$ is the reference image reconstructed with common LO and $I_{Test}$ is the image reconstructed using either the USBS-derived LOs or the two independent free-running LOs. The common-LO configuration represents the reference case in which synchronization-induced differential LO phase noise is eliminated while the remaining receiver, digitizer, reconstruction, and geometric errors are retained.

At the emitter location, this degradation is 37.5 dB with two independent free-running LOs. Because the LO-drive power varies by approximately 2 dB across the three test cases, leading to corresponding differences in intermediate frequency (IF) voltages sampled by the oscilloscope, part of the observed magnitude difference may therefore be attributed to differences in the hardware implementation. Consequently, the measured 1.7 dB image-magnitude difference should not be interpreted as purely synchronization loss.

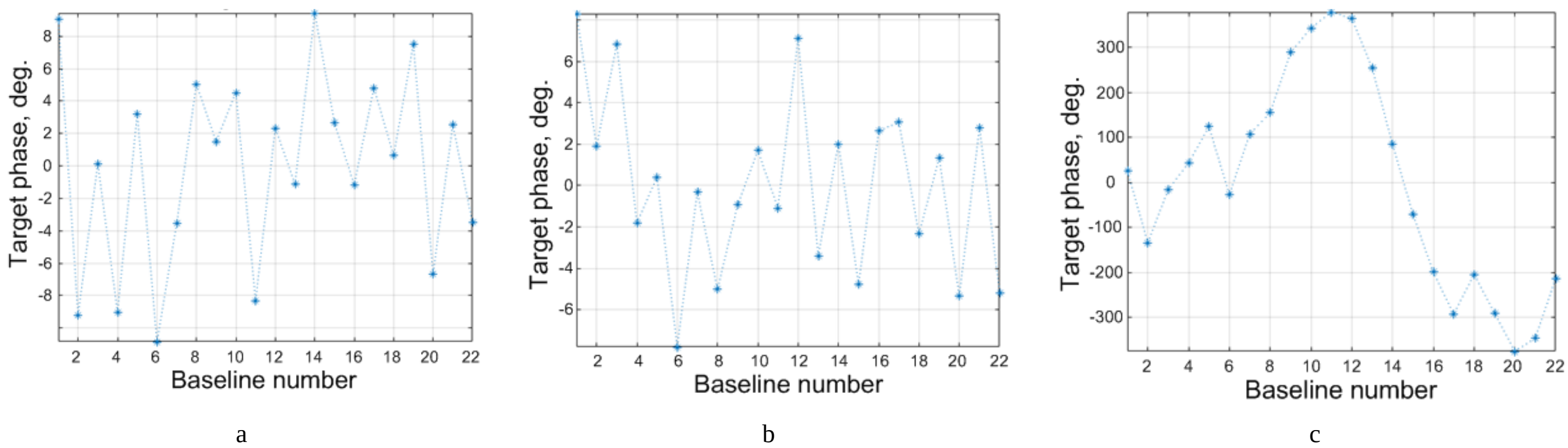

Fig. 9. Unwrapped phase as a function of antenna position using (a) the common LO, (b) LOs under test, (c) free-running LOs. Phases are evaluated at the emitter location. Each realization is centered on its mean for visualization

We define the target-pixel coherence at the emitter pixel as

$$C = \frac{|\sum_a z|}{\sum_a |z|}, \tag{4}$$

where index $a$ denotes the antenna/baseline position and $z = I_a(\boldsymbol{r_0})$ is the image value at the emitter location. The resulting coherence values are summarized in Table 2.

The residual phase standard deviation is measured at the emitter pixel.

The W-band interferometric performance of P=2 USBS is summarized in Table 2. The coherence factor C=0.985 corresponds to only ~0.13 dB coherent-amplitude loss; therefore, most of the observed 1.7 dB difference is attributed to signal-chain amplitude differences rather than synchronization-induced decoherence. At free-running LO setup, the cross-correlation degrades towards the residual floor (approximately $10^{-3}$ at the Table 2) on millisecond timescales and does not recover at longer integration times.

Table 2 – Summary

| | Common LO | USBS | Free-Running |
|---|---|---|---|
| **Cross-correlation magnitude at $T = 500$ s** | *0.99* | *0.98* | *residual floor ~$10^{-3}$* |
| **Target-pixel coherence $C$** | *0.995* | *0.985* | *0.22* |
| **Residual phase standard deviation** | *9.4°* | *13.1°* | *175.4°* |
| **Image magnitude loss $L$, dB** | *Reference* | *1.7 dB* | *37.5 dB* |

The experiment was designed to isolate synchronization-induced coherence degradation. It therefore does not reproduce all error sources of a flight interferometer, including independent receiver thermal variations, astronomical calibration etc. Furthermore, the 500 s effective integration comprises multiple short acquisitions distributed over approximately five days rather than a continuous 500 s acquisition. The results should therefore be interpreted as an instrument-level validation of the synchronization concept under controlled laboratory conditions.

## 5. Conclusions

This work presented the first instrument-level validation of the Upper Sideband Syntonization (USBS) concept. Unlike previous investigations, which characterized the concept primarily through frequency stability and phase-difference measurements, the present work evaluated its impact on instrument-level performance through interferometric observables and image reconstruction.

Experimental results obtained with a W-band radio interferometer demonstrate that local oscillators (LOs) derived from the USBS P=2 breadboard preserve a coherence of 0.98 in both long-term correlation measurements and image reconstruction. The demonstrated capability to maintain coherent correlation over 500 s of cumulative effective integration time acquired over approximately five days, together with the observed inter-session phase stability over a 40-day measurement span, indicates synchronization performance relevant to the coherent-integration requirements of space-based Event Horizon Imaging. The demonstrated phase stability may enable novel space-interferometry mission concepts and support the development of new phase-calibration strategies.

The image reconstructed using LOs derived from the P=2 USBS breadboard is comparable to that obtained with a conventional common-LO architecture, whereas operation with two independent free-running LOs leads to severe coherence degradation down to 0.22, a 37.5 dB magnitude reduction at the true emitter location, together with displacement and defocusing of the reconstructed response. The image-validation results therefore demonstrate that USBS enables coherent image reconstruction.

The reported results provide instrument-level validation of the USBS concept and demonstrate that its performance is sufficient for coherent operation of distributed radio interferometric receivers under the tested laboratory conditions. This constitutes an important step toward practical implementation of USBS in future distributed space interferometers.

### Acknowledgments

The authors acknowledge N. Ayllon, E. Lia, M. Miller, and E. van der Houwen of D/TEC, ESA/ESTEC, for their substantial support with the microwave hardware, including modifications to the RADET receivers and provision of the Thorlabs translation stages and IF amplifiers. The first author thanks his family for their support during the completion of this work.

### ORCID

Volodymyr Kudriashov - https://orcid.org/0000-0001-6503-3063

Manuel Martin-Neira - https://orcid.org/0000-0001-7626-7033